\documentclass[prd,aps,twocolumn,amsmath,amssymb,nofootinbib,preprintnumbers]
{revtex4}

\usepackage{graphicx}
\usepackage{dcolumn}
\usepackage{bm}
\usepackage{amsmath}
\usepackage{amsthm}
\usepackage{amsfonts}
\usepackage{euscript,bbm}
\usepackage{ifthen}
\usepackage{psfrag}
\usepackage{slashed}

\def\ls{\mathrel{\lower4pt\vbox{\lineskip=0pt\baselineskip=0pt
           \hbox{$<$}\hbox{$\sim$}}}}
\def\gs{\mathrel{\lower4pt\vbox{\lineskip=0pt\baselineskip=0pt
           \hbox{$>$}\hbox{$\sim$}}}}
\def\drawbox#1#2{\hrule height#2pt
\hbox{\vrule width#2pt height#1pt \kern#1pt
              \vrule width#2pt}
              \hrule height#2pt}

\def\Asym#1#2{\vcenter{\vbox{\drawbox{#1}{#2}
              \kern-#2pt       
              \drawbox{#1}{#2}}}}

\newcommand{\be}{\begin{equation}}
\newcommand{\ee}{\end{equation}}
\newcommand{\bea}{\begin{eqnarray}}
\newcommand{\eea}{\end{eqnarray}}

\newcommand{\neu}[1]{\ensuremath{\tilde{\chi}_{#1}^0}}
\newcommand{\chp}[1]{\ensuremath{\tilde{\chi}_{#1}^+}}
\newcommand{\chm}[1]{\ensuremath{\tilde{\chi}_{#1}^-}}

\newcommand{\chpm}[1]{\ensuremath{\tilde{\chi}_{#1}^{\pm}}}
\newcommand{\chmp}[1]{\ensuremath{\tilde{\chi}_{#1}^{\mp}}}

\newcommand{\st}{\ensuremath{\tilde{t}}}

\newcommand{\gsim}{\lower.7ex\hbox{$\;\stackrel{\textstyle>}{\sim}\;$}}
\newcommand{\lsim}{\lower.7ex\hbox{$\;\stackrel{\textstyle<}{\sim}\;$}}
\newcommand{\tbar}{\overline{t}}

\newcommand{\met}{{E\!\!\!\!/_{\rm T}}}

\newcommand{\pT}{{p_{\rm T}}}

\newcommand{\pythia}{{\tt PYTHIA}}

\newcommand{\pgs}{{\tt PGS4}}

\newcommand{\madgraph}{{\tt MADGRAPH5}}

\newcommand{\ben}{\begin{enumerate}}
\newcommand{\een}{\end{enumerate}}
\newcommand{\bei}{\begin{itemize}}
\newcommand{\eei}{\end{itemize}}

\begin{document}

\title{Probing Supersymmetric Dark Matter and the Electroweak Sector using Vector Boson Fusion Processes: A Snowmass Whitepaper}

\author{Andres G. Delannoy$^{2}$}
\author{Bhaskar Dutta$^{1}$}
\author{Alfredo Gurrola$^{2}$}
\author{Will Johns$^{2}$}
\author{Teruki Kamon$^{1,3}$}
\author{Eduardo Luiggi$^{4}$}
\author{Andrew Melo$^{2}$}
\author{Paul Sheldon$^{2}$}
\author{Kuver Sinha$^{5}$}
\author{Kechen Wang$^{1}$}
\author{Sean Wu$^{1}$}

\affiliation{$^{1}$~Mitchell Institute for Fundamental Physics and Astronomy, \\
Department of Physics and Astronomy, Texas A\&M University, College Station, TX 77843-4242, USA \\
$^{2}$~Department of Physics and Astronomy, Vanderbilt University, Nashville, TN, 37235, USA \\
$^{3}$~Department of Physics, Kyungpook National University, Daegu 702-701, South Korea \\
$^{4}$~Department of Physics, University of Colorado, Boulder, CO 80309-0390, USA \\
$^{5}$~Department of Physics, Syracuse University, Syracuse, NY 13244, USA
}

\begin{abstract}

Vector boson fusion (VBF) processes at the Large Hadron Collider (LHC) provide a unique opportunity to search for new physics with electroweak couplings. Two studies are presented: $(i)$ A search of supersymmetric dark matter in the final state of two VBF jets and large missing transverse energy is presented at 14 TeV. Prospects for determining the dark matter relic density are studied for the cases of Wino and Bino-Higgsino dark matter. The LHC could probe Wino dark matter with mass up to approximately 600 GeV with a luminosity of 1000 fb$^{-1}$. $(ii)$ A search for the chargino/neutralino system in the final state of two VBF jets, missing transverse energy and two $\tau$s (light stau case) and light lepton $e$ and $\mu$ (light slepton case). The $5 \sigma$ mass reach at $300$ fb$^{-1}$ ($1000$ fb$^{-1}$) of LHC14 for inclusive and opposite-sign $\tau$ pairs are $250$ GeV ($300$ GeV) and $200$ GeV ($250$ GeV), respectively, for $\Delta M \, = \, m_{\tilde{\tau}_1} - m_{\neu{1}} \, = \, 30$ GeV. For $\Delta M \, = \, 15$ GeV, the $3 \sigma$ mass reach at $300$ fb$^{-1}$ ($1000$ fb$^{-1}$) of LHC14 for inclusive $\tau$ pairs is $180$ GeV. The $5 \sigma$ mass reach at $300$ fb$^{-1}$ ($1000$ fb$^{-1}$) of LHC14 for inclusive and opposite-sign $\mu$ pairs are approximately $350$ GeV ($400$ GeV) and $300$ GeV ($350$ GeV), respectively. The mass reaches in the same-sign final state cases are similar to those in the opposite-sign cases.

\end{abstract}

\maketitle

\section{Introduction}

Recently, experiments at the 8-TeV LHC (LHC8) have put lower bounds on the masses of the $\tilde{g}$ and $\tilde{q}$. For comparable masses, the exclusion limits are approximately $1.5$ TeV at $95\%$ CL with $13$ fb$^{-1}$ of integrated luminosity \cite{LHCsquarkgluino}. For $20$ fb$^{-1}$ analysis, see \cite{LHCsquarkgluino20ifb}. There has also been active investigation (both theoretical and experimental) for the lightest top squark ($\tilde{t}$), and exclusion limits in the $m_{\st}$-$m_{\neu{1}}$ plane have been obtained in certain decay modes  \cite{stops}. 

 A variety of possibilities exist for the colored sector (compressed spectra, mildly fine-tuned split scenarios \cite{squarkheavy}, non-minimal supersymmetric extensions, etc.) with varying implications for existing and future searches. From the perspective of a hadron collider, where electroweak (EW) production is small, a classic strategy to study the chargino/neutralino system is to detect the neutralinos in cascade decays of gluinos and squarks. For example, reconstructing a decay chain like $\tilde{g} \, \rightarrow \, \tilde{q} \, \rightarrow \, \neu{2} \, \rightarrow \, \tilde{\tau}_1 \, \rightarrow \, \neu{1}$ using endpoint methods \cite{Arnowitt:2008bz, Hinchliffe:1996iu} leads to  mass measurements of $\neu{2}$, $\tilde\tau_1$, $\tilde q$ and $\tilde g$,  where $\tilde\tau_1$ is the lighter stau mass \cite{Dutta:2011kp}. However, in a scenario where colored objects are heavy and the production cross-section is limited, one has to use different techniques to probe the EW sector. Moreover, experimental constraints (e.g. triggering) significantly affect the ability to probe supersymmetric EW sector in some of the above scenarios. 
 
The important point to note is that \textit{a direct probe of the EW sector is largely agnostic about the fate of the colored sector} and provides a window to dark matter (DM) physics. Bounds on directly produced charginos and neutralinos in final states with three leptons and missing transverse momentum using $20.7$ fb$^{−1}$ of integrated luminosity at LHC8 have been presented by the ATLAS collaboration in \cite{ATLASneutchargino}. Similarly, the CMS collaboration has investigated EW production of charginos, neutralinos, and sleptons in final states with exactly three leptons, four leptons, two same-sign leptons, two opposite-sign-same-flavor leptons plus two jets, and two opposite-sign leptons inconsistent with $Z$ boson decay, at an integrated luminosity of $9.2$ fb$^{-1}$ at LHC8 \cite{CMSneutchargino}.

The purpose of this whitepaper is to summarize work recently done by the authors in \cite{Dutta:2012xe} and \cite{Delannoy:2013ata}, where the EW sector has been investigated using vector boson fusion (VBF) processes \cite{Cahn:1983ip, Bjorken:1992er}. VBF processes have been suggested for Higgs searches \cite{Rainwater:1998kj} and supersymmetric searches, in the context of slepton and gaugino productions at $14$ TeV LHC \cite{Choudhury:2003hq,cho,datta}. 


The whitepaper is divided into two parts, dedicated to the two separate studies:

$(i)$ Direct DM production by VBF processes in events with $2j \, + \, \met$ in the final state. Information about production cross sections in VBF processes and the distribution of $\met$ in the final state can be used to solve for the mass and composition of $\neu{1}$, and hence the DM relic density. The cases of pure Wino or Higgsino $\neu{1}$, as well as the case of a mixed Bino-Higgsino $\neu{1}$ are studied. 

$(ii)$ Probing $\neu{2}$, $\chpm{1}$ with VBF processes. We note that the analysis in \cite{Dutta:2012xe} was presented for LHC8, and work is currently being done to upgrade our studies to the 14 TeV LHC (LHC14).

\section{Probing Dark Matter at the LHC using VBF Processes}

Nearly $80\%$ of the matter of the Universe is dark matter (DM) \cite{WMAP}. The identity of DM is one of the most profound questions at the interface of particle physics and cosmology. Weakly interacting massive particles (WIMPs) are particularly promising DM candidates that can explain the observed relic density and are under investigation in a variety of direct and indirect searches. Within the context of $R$-parity conserving supersymmetric extensions of the standard model (SM), the WIMP DM candidate is the lightest supersymmetric particle (LSP), typically the lightest neutralino ($\neu{1}$), which is a mixture of Bino, Wino, and Higgsino states. 

The DM relic density is typically determined by its annihilation cross section at the time of thermal freeze-out. For supersymmetric WIMP DM, the annihilation cross section depends on the mass of $\neu{1}$ and its couplings to various SM final states, for which a detailed knowledge of the composition of $\neu{1}$ in gaugino/Higgsino states is required. Moreover, other states in the electroweak sector, such as sleptons, staus, or charginos can enter the relic density calculation.

The strategy pursued in this section will be to investigate direct DM production by VBF processes in events with $2j \, + \, \met$ in the final state. Such an approach has several advantages. The $2j \, + \, \met$ final state configuration provides a search strategy that is free from trigger bias. This is reinforced as the $p_T$ thresholds for triggering objects are raised by ATLAS and CMS experiments.

In order to probe DM directly, the following processes are investigated:
%
%
\be
pp \rightarrow \neu{1} \, \neu{1} \, jj, \,\,\, \chpm{1} \, \chmp{1} \, jj , \,\,\, \chpm{1} \, \neu{1} \, jj \,\,\,\,.
\ee
The main sources of SM background are: $(i)$ \,\, $pp \rightarrow Z jj \rightarrow \nu \nu j j $ and $(ii)$ \,\, $pp \rightarrow W jj \rightarrow l \nu j j $. The former is an irreducible background with the same topology as the signal. The $\met$ comes from the neutrinos. The latter arises from events which survive a lepton veto; $(iii)$ \,\, $pp \rightarrow t\tbar + $jets: This background may be reduced by vetoing $b$-jets, light leptons, $\tau$ leptons and light-quark/gluon jets.

The search strategy relies on requiring the tagged VBF jets, vetoes for $b$-jets, light leptons, $\tau$ leptons and light-quark/gluon jets, and requiring large $\met$ in the event. Signal and background events are generated with \madgraph \,\, \cite{Alwall:2011uj}. The \madgraph \, events are then passed through \pythia\,\cite{pythia} for parton showering and hadronization. The detector simulation code used here in this work is \pgs \, \cite{pgs}. 

Distributions of $\pT(j_1), \pT(j_2), M_{j_1j_2}$, and $\met$ for background as well as VBF pair production of DM are studied at $\sqrt{s}=8$ TeV and 14 TeV. In the case of pure Wino or Higgsino DM, $\chpm{1}$ is taken to be outside the exclusion limits for ATLAS' disappearing track analysis \cite{chargedtrack} and thus VBF production of $\chpm{1} \chpm{1}$, $\chpm{1} \chmp{1}$, and $\chpm{1} \neu{1}$ also contribute. The $\neu{1}$ masses chosen for this study are in the range 100 GeV to 1 TeV. The colored sector is assumed to be much heavier. There is no contribution to the neutralino production from cascade decays of colored particles. 

Events are preselected by requiring $\met > 50$ GeV and the two leading jets ($j_{1}$,$j_{2}$) each satisfying $p_{T} \geq 30$ GeV with $|\Delta \eta(j_{1},j_{2})| > 4.2$ and $\eta_{j_{1}}\eta_{j_{2}} < 0$. The preselected events are used to optimize the final selections to achieve maximal signal significance ($S / \sqrt{S + B}$). For the final selections, the following cuts are employed: $(i)$ The tagged jets are required to have $p_{T} > 50$ GeV and $M_{j_{1} j_{2}} > 1500$ GeV; $(ii)$ Events with loosely identified leptons ($l = e,\mu,\tau_{h}$) and $b$-quark jets are rejected, reducing the $t \tbar$ and $Wjj \rightarrow l\nu jj$ backgrounds by approximately $10^{-2}$ and $10^{-1}$, respectively, while achieving $99\%$ efficiency for signal events. The $b$-jet tagging efficiency used in this study is $70\%$ with a misidentification probability of $1.5\%$, following Ref. \cite{Chatrchyan:2012jua}. Events with a third jet (with $p_{T} > 30$ GeV) residing between $\eta_{j_{1}}$ and $\eta_{j_{2}}$ are also rejected; $(iii)$ The $\met$ cut is optimized for each different value of the DM mass. For $m_{\neu{1}} = 100$ GeV ($1$ TeV), $\met \geq 200$ GeV ($450$ GeV) is chosen, reducing the $Wjj\rightarrow l\nu jj$ background by approximately $10^{-3} \, (10^{-4})$. 
We have checked and found that missing energy is the biggest discriminator between background and signal events. After the missing energy cut, the azimuthal angle difference of the two tagging jets~\cite{zeppenfeld} does not improve the search limit.

\begin{figure}[!htp]
\centering
\includegraphics[width=3.5in]{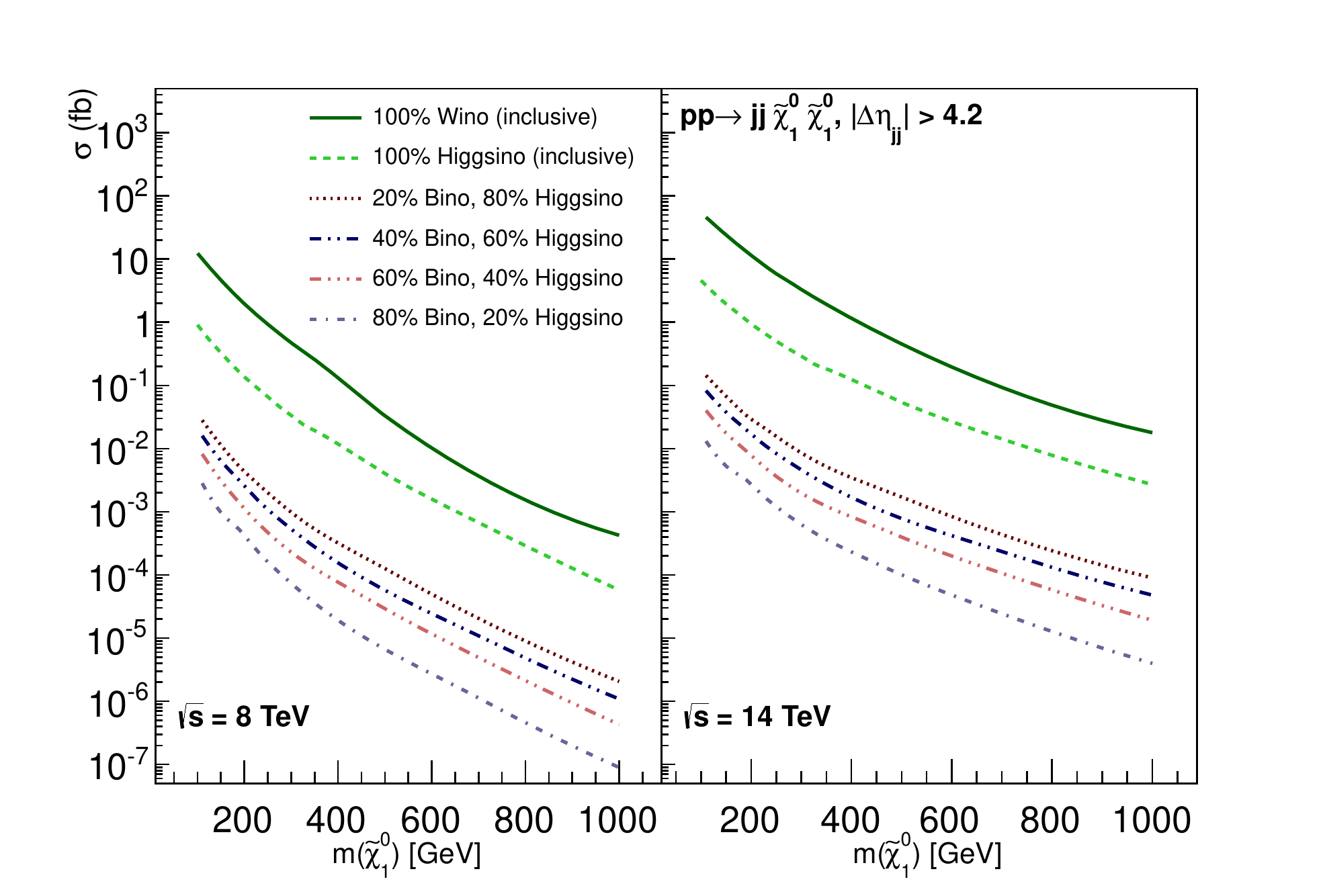}
\caption{Production cross section as a function of $m_{\neu{1}}$ after requiring $|\Delta \eta (j_1, j_2)| > 4.2$, at LHC8 and LHC14. For the pure Wino and Higgsino cases, inclusive $\neu{1} \neu{1}$,  $\chpm{1} \chpm{1}$, $\chpm{1} \chmp{1}$, and $\chpm{1} \neu{1}$ production cross sections are displayed.}
\label{xsectionColor}
\end{figure}

The production cross section as a function of $m_{\neu{1}}$ after requiring $|\Delta \eta (j_1, j_2)| > 4.2$ is displayed in Fig. \ref{xsectionColor}. The left and right panels show the cross sections for LHC8 and LHC14, respectively. For the pure Wino and Higgsino cases, inclusive $\neu{1} \neu{1}$,  $\chpm{1} \chpm{1}$, $\chpm{1} \chmp{1}$, and $\chpm{1} \neu{1}$ production cross sections are displayed. The green (solid) curve corresponds to the case where $\neu{1}$ is $99\%$ Wino. The inclusive production cross section is $\sim 40$ fb for a $100$ GeV Wino at LHC14, and falls steadily with increasing mass. The cross section is approximately $5-10$ times smaller for the pure Higgsino case, represented by the green (dashed) curve. As the Higgsino fraction in $\neu{1}$ decreases for a given mass, the cross section drops. For $20\%$ Higgsino fraction in $\neu{1}$, the cross section is $ \sim 10^{-2}$ fb for $m_{\neu{1}} = 100$ GeV at LHC14.

\begin{figure}[!htp]
\centering
\includegraphics[width=3.5in]{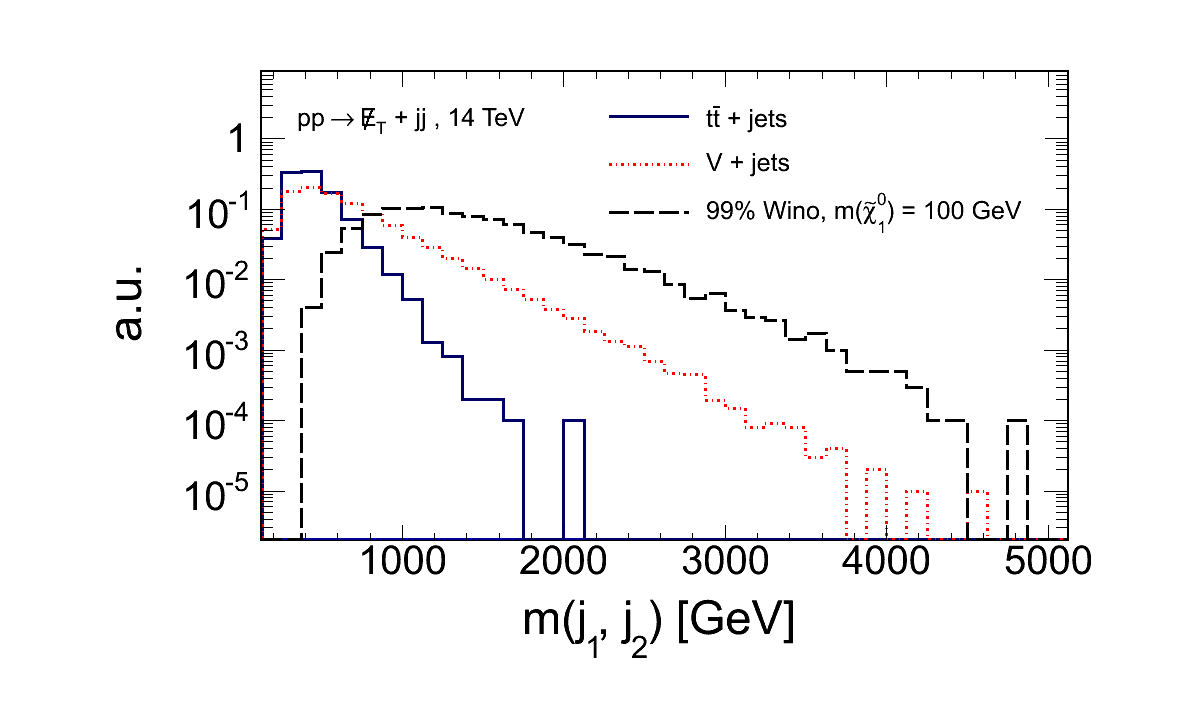}
\caption{Distribution of the dijet invariant mass $M_{j_1j_2}$ normalized to unity for the tagging jet pair $(j_1,j_2)$ and main sources of background after pre-selection cuts and requiring $p_T > 50$ GeV for the tagging jets at LHC14. The dashed black curves show the distribution for the case where $\neu{1}$ is a nearly pure Wino with $m_{\neu{1}} = 50$ and $100$ GeV. Inclusive $\neu{1} \neu{1}$,  $\chpm{1} \chpm{1}$, $\chpm{1} \chmp{1}$, and $\chpm{1} \neu{1}$ production is considered.}
\label{DiJetMass_VBFDM}
\end{figure}

Figure \ref{DiJetMass_VBFDM} shows the dijet invariant mass distribution $M_{j_1j_2}$ for the tagging jet pair $(j_1,j_2)$ and main sources of background, after the pre-selection cuts and requiring $p_T > 50$ GeV for the tagging jets at LHC14. The dashed black curves show the distribution for the case of a pure Wino DM, with $m_{\neu{1}} = 50$ and $100$ GeV.  The dijet invariant mass distribution for $W+$ jets, $Z+$ jets, and $t \bar{t} +$ jets background are also displayed. Clearly, requiring $M_{j_1j_2} > 1500$ GeV is effective in rejecting background events, resulting in a reduction rate between $10^{-4}$ and $10^{-2}$ for the backgrounds of interest.

\begin{figure}[!htp]
\centering
\includegraphics[width=3.5in]{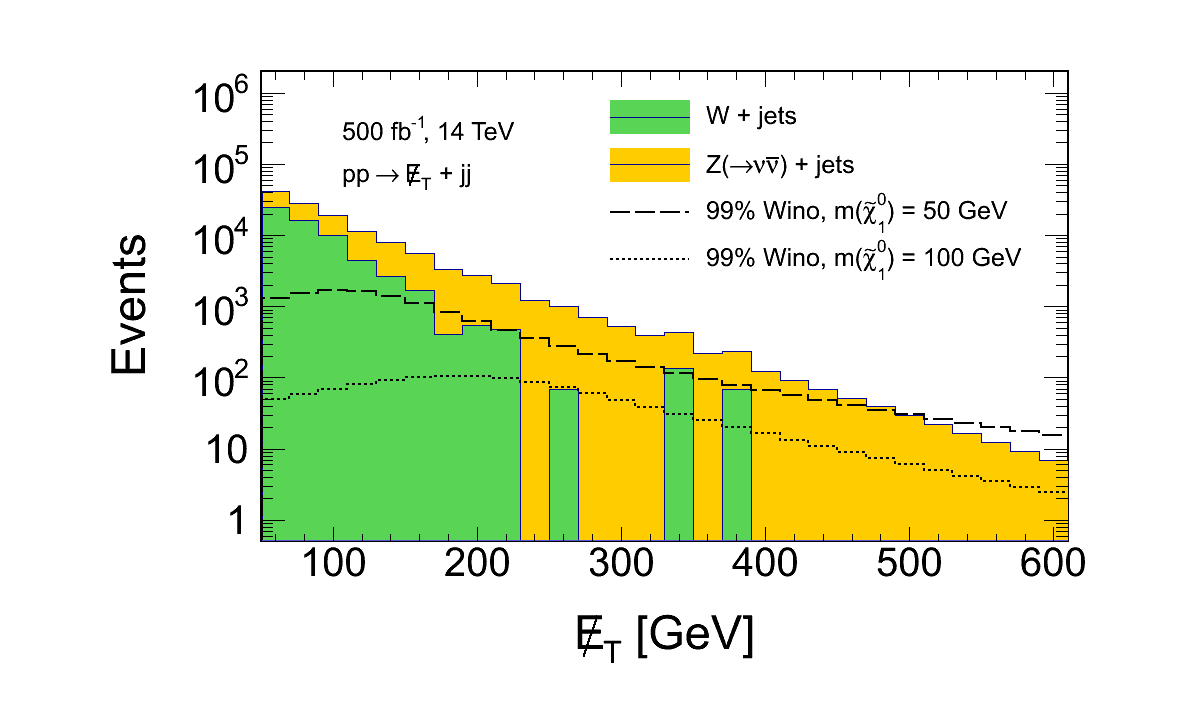}
\caption{The $\met$ distributions for Wino DM (50 GeV and 100 GeV) compared to $W+$ jets and $Z+$ jets events with $500$ fb$^{-1}$ integrated luminosity at LHC14. The distributions are after all selections except the $\met$ cut. Inclusive $\neu{1} \neu{1}$,  $\chpm{1} \chpm{1}$, $\chpm{1} \chmp{1}$, and $\chpm{1} \neu{1}$ production is considered.}
\label{Met_VBFDM_versionB}
\end{figure}

Figure \ref{Met_VBFDM_versionB} shows the $\met$ distribution for an integrated luminosity of 500 fb$^{-1}$ at LHC14  after all final selections except the $\met$ requirement. There is a significant enhancement of signal events in the high $\met$ region.


\begin{figure}[!htp]
\centering
\includegraphics[width=3.5in]{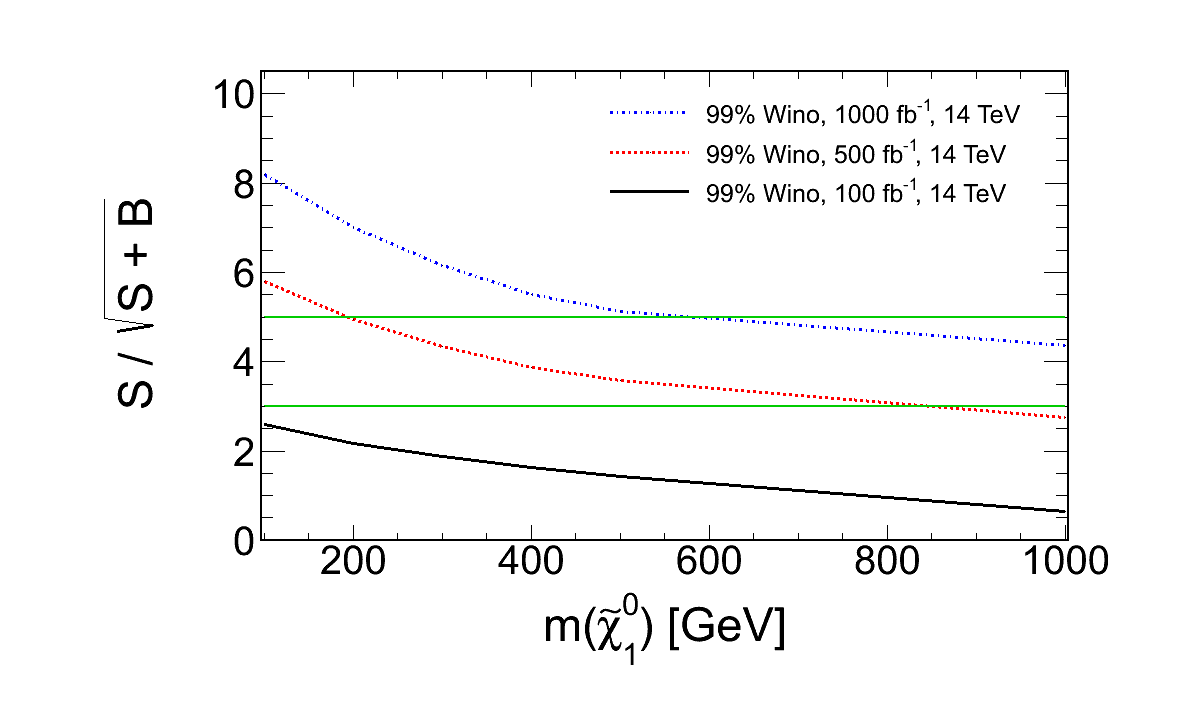}
\caption{Significance curves for the case where $\neu{1}$ is $99\%$ Wino as a function of $m_{\neu{1}}$ mass for different luminosities at LHC14. The green lines correspond to $3\sigma$ and $5\sigma$ significances.}
\label{SignificanceVsMassVsLumi_versionD}
\end{figure}

The significance as a function of $\neu{1}$ mass is plotted in Fig. \ref{SignificanceVsMassVsLumi_versionD} for different luminosities at LHC14. 
The blue, red, and black curves correspond to luminosities of $1000, 500,$ and $100$ fb$^{-1}$, respectively. At $1000$ fb$^{-1}$, a significance of $5\sigma$ can be obtained up to a Wino mass of approximately $600$ GeV.  
The analysis is repeated by changing the jet energy scale and lepton energy scale by 20\% and 5\%, respectively.
We find the uncertainties in the significance to be 4\%.

Determining the composition of $\neu{1}$ for a given mass is very important in order to understand early universe cosmology. For example, if $\neu{1}$ has a large Higgsino or Wino component, the annihilation cross section is too large to fit the observed relic density for $m_{\neu{1}}$ mass less than $\sim 1$ TeV for Higgsinos \cite{Allahverdi:2012wb} and $\sim 2.5$ TeV for Winos. On the other hand if $\neu{1}$ is mostly Bino, the annihilation cross section is too small. In the first case one has under-abundance whereas in the second case one has over-abundance of DM. Both problems can be solved if the DM is non-thermal \cite{Allahverdi:2012gk} (in the case of thermal DM, addressing the over abundance  problem requires addition effects like resonance, coannihilation etc. in the cross section, while the under-abundance problem can be addressed by having multi-component DM \cite{Baer:2012cf}). If  $\neu{1}$ is a suitable mixture of Bino and Higgsino, the observed DM relic density can be satisfied.   

From Figs. \ref{xsectionColor} and \ref{Met_VBFDM_versionB}, it is clear that varying of the rate and the shape of the $\met$ distribution can be used to solve for the mass of $\neu{1}$ as well as its composition in gaugino/Higgsino eigenstates. The VBF study described in this work was performed over a grid of input points on the $F - m_{\neu{1}}$ plane (where $F$ is the Wino or Higgsino percentage in $\neu{1}$). The $\met$ cut was optimized over the grid, and the $\met$ shape and observed rate of data were used to extract $F$ and $m_{\neu{1}}$ which was then used to determine the DM relic density. 



\begin{figure}[!htp]
\centering
\includegraphics[width=3.5in]{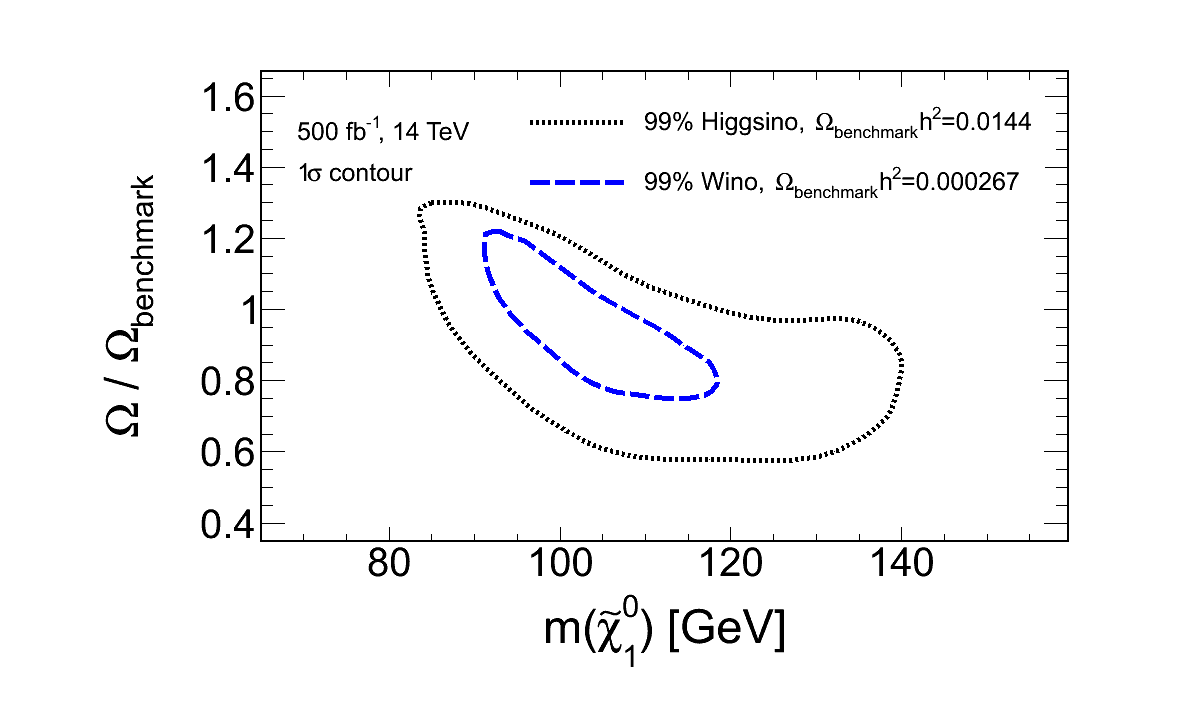}
\caption{Contour lines in the relic density-$m_{\neu{1}}$ plane for $99\%$ Wino (blue dashed) and $99\%$ Higgsino (grey dotted) DMs expected with $500$ fb$^{-1}$ of luminosity at LHC14. The relic density is normalized to its value at $m_{\neu{1}} = 100$ GeV.}
\label{OmegaVsMLSP_WinoAndHiggsino_versionD}
\end{figure}

In Fig. \ref{OmegaVsMLSP_WinoAndHiggsino_versionD}, the case of $99\%$ Higgsino and $99\%$ Wino were chosen, and $1\sigma$ contour plots drawn on the relic density-$m_{\neu{1}}$ plane for $500$ fb$^{-1}$ luminosity at LHC14. The relic density was normalized to a benchmark value $\Omega_{\rm benchmark}$, which is the relic density for $m_{\neu{1}} = 100$ GeV.  For the Wino case, the relic density can be determined within $\sim 20\%$, while for the Higgsino case it can be determined within $\sim 40\%$. For higher values of $m_{\neu{1}}$, higher luminosities would be required to achieve these results. We note we have not evaluated the impact of any degradation in $\met$ scale, linearity and resolution due to large pile-up events. Our results represent the best case scenario and it will be crucial to revisit with the expected performance of upgraded ATLAS and CMS detectors. 

\section{Probing $\neu{2}$, $\chpm{1}$ with VBF Processes}

In this section, we probe $\neu{2}$, $\chpm{1}$ with VBF processes in final states with  $2 \tau \, + \, \met$ as well as $2 l \, + \, \met$, based on \cite{Dutta:2012xe}. 

We note that the analysis in \cite{Dutta:2012xe} was presented at the 8 TeV LHC, and work is currently being done to upgrade our studies to 14 TeV. Much of the VBF kinematics of the $\neu{2}$, $\chpm{1}$ study at 14 TeV will be similar to the 14 TeV analysis of DM presented in the previous section. We thus present the 8 TeV analysis below, with final significances at 14 TeV obtained by a simple scaling from the 8 TeV results.

Several features of existing searches for the $\neu{2}$, $\chpm{1}$ system at the LHC may be highlighted for better contrast with the present study:

$(a)$ Searches target $\neu{2}$ and $\chpm{1}$ produced via Drell-Yan production, i.e., without requiring the presence of forward/backward jets.

$(b)$ The bounds assume $m_{\tilde{\chi}^{\pm}_1} \, \sim \, m_{\neu{2}}$ and an enhanced branching ratio to trilepton final state provided by $m_{\tilde{l}_1} \sim (m_{\tilde{\chi}^{\pm}_1} + m_{\neu{1}})/2$ (where $\tilde{l}_1$ is the lighter slepton, and $l$ denotes either $e$ or $\mu$). The slepton mass $m_{\tilde{l}_1}$ is placed such that the mass splitting is large and thus the leptons have relatively high $p_{T}$ to be free from trigger bias.

$(c)$ Searches resulting in $\tau$ final states do not exist due to the larger $\tau$ misidentification rates which make it difficult to both manage the level of backgrounds and maintain low enough $p_{T}$ thresholds for triggering.

Probing $\neu{2}$, $\chpm{1}$ with VBF processes offers certain advantages:

$(i)$ With increasing instantaneous luminosity at the LHC, both ATLAS and CMS experiments are raising their $\pT$ thresholds for triggering any object. This motivates us to probe signals for supersymmetry in VBF processes where production of superpartners is free from trigger bias.

$(ii)$ VBF production allows the investigation of final states with $\tau$. $\tilde\tau_1$  is typically lighter than $\tilde\mu_1$ and $\tilde e_1$ for large $\tan\beta$. A light $\tilde{\tau}_1$ with small mass splitting is favored in coannihilation processes \cite{Griest:1990kh} that set the relic density to correct values, in the case of Bino dark matter. Light $\tilde{\tau}_1$ is also motivated in the context of the MSSM by the enhancement of the $h \, \rightarrow \, \gamma \gamma$ channel \cite{Carena:2011aa}. These facts stress the importance of searches in $\tau$ final states with low $\pT$ and large backgrounds, for which production by VBF processes is more suited since the VBF signature allows for the reduction of the backgrounds to manageable levels.

$(iii)$ For the leptonic final state, a search based on VBF processes can be complementary or better than the existing LHC searches based on Drell-Yan production, since it is not constrained by trigger bias. It is also interesting to note that the Drell-Yan production cross-section falls faster than the VBF production cross-section with increasing mass \cite{Choudhury:2003hq}.

We now present the main results of the analysis.

The $\neu{2}$ and $\chpm{1}$ are produced by VBF processes and then decay into  the lighter slepton states ($\tilde\tau_1$,  $\tilde\mu_1$ and  $\tilde e_1$) by the decay processes $\tilde{\chi}^{\pm}_1 \, \rightarrow \, \tilde{\tau}_1 \nu \, \rightarrow \, \neu{1} \tau \nu$, $\tilde{\chi}^{\pm}_1 \, \rightarrow \, \tilde{ l} \nu \, \rightarrow \, \neu{1} l \nu$, and similarly for $\neu{2}$ via $\neu{2}\, \rightarrow \, \tilde{\tau}_1 \tau \, \rightarrow \, \neu{1} \tau \tau$ and $\neu{2}\, \rightarrow \, \tilde{l}_1 l \, \rightarrow \, \neu{1} ll$.

A benchmark point is first defined and the following processes are investigated:
\be
pp \rightarrow \chpm{1} \, \chpm{1} \, jj , \,\,\, \chp{1} \, \chm{1} \, jj , \,\,\, \chpm{1} \, \neu{2} jj ,\,\,\, \neu{2} \, \neu{2} jj \,\,.
\ee
The benchmark point is $m_{\tilde{\chi}^{\pm}_1} \, \sim \, m_{\tilde{\chi}^0_2} \, = 181$ GeV, $m_{\tilde{\tau}_1} = 130$ GeV, and $m_{\tilde{\chi}^0_1} = 100$ GeV. The $\chpm{1}$ and \neu{2} are mainly Wino, while $\neu{1}$ is mainly Bino. 

The search strategy is based on two steps: first, use the unique features of VBF processes to reduce background $V+$jets events (where $V$ is either $W$ or $Z$), and second, use decay properties of the centrally produced supersymmetric particle to reduce non-supersymmetric channels that are also produced by VBF processes. 

The production of  $VV$ (where $V$ may be wither $W$ or $Z$) by VBF processes mimics the signal when the bosons decay leptonically. A $\met$ cut is effective in reducing this background. 
Moreover, requiring multiple $\tau$'s in the event further reduces background. Results will be presented for requiring same-sign and oppositely-signed $\tau$ pairs, as well as an inclusive study. Although  $m_{\tilde{\chi}^{\pm}_1} \, \sim \, m_{\tilde{\chi}^0_2}$ is chosen as an example, the methods described in this paper are applicable in detecting $\neu{2}$ and $\chpm{1}$ separately.

Signal and background samples are generated with \madgraph \,\, \cite{Alwall:2011uj} followed by detector simulation using \pgs \, \cite{pgs}.

For the VBF selections, we accept jets with $\pT \geq 50$ GeV in $|\eta| \leq 5$, and require a presence of two jets ($j_1$, $j_2$) satisfying: 

$(i)$ $\pT(j_1)  \geq 75$ GeV;

$(ii)$ $|\Delta \eta (j_1, j_2)| > 4.2$;

$(iii)$ $\eta_{j_1} \eta_{j_2} < 0$;

$(iv)$ $M_{j_1j_2} > 650$ GeV.

We note that the signal acceptance with this selection  is less sensitive to effects on the signal acceptance due to initial/final state radiation, pileup, and fluctuations in jet fragmentation.

Although the central jet veto has been used in the past for VBF Higgs, we
do not employ a central veto
cut in our case as our backgrounds are already small and such a veto in
high pileup conditions can degrade
the signal acceptance and requires extensive study in the future.

With our proposed  VBF cuts, it is fruitful to divide the study into the $2 \, \tau \, + \met$ and $2 \, l \, + \met$ final states separately.

\subsection{$\geq 2 j \, + \, 2 \tau \, + \, \met$}

For this final state, the following selections are employed in addition to the VBF cuts described above:

$(i)$ Two $\tau$'s with $\pT \geq 20$ GeV in $|\eta| < 2.1$, with $\Delta R (\tau, \tau) > 0.3$. All $\tau$'s considered in this paper are hadronic. The $\tau$ ID efficiency is assumed to be $55\%$ and the jet $\to \tau$ misidentification rate is taken to be $1\%$, both flat over $\pT$ \cite{cmstau}. 
A branching ratio of $100\%$ of $\chpm{1}$ and $\neu{2}$ to $\tilde{\tau}_1$ has been assumed. In realistic models, this branching ratio can be close to  $100\%$.


$(ii)$ $\met > 75$ GeV. This cut is expected to be effective, due to the fact that the main source of $\met$ for signal is the $\neu{1}$ LSP which leaves the detector, while for the background it is the neutrinos produced by leptonic decays of the vector bosons. 

$(iii)$ We also apply a loose $b$-veto which is useful in reducing the $t\overline{t}$ background.

\begin{figure}[!htp]
\centering
\includegraphics[width=3.5in]{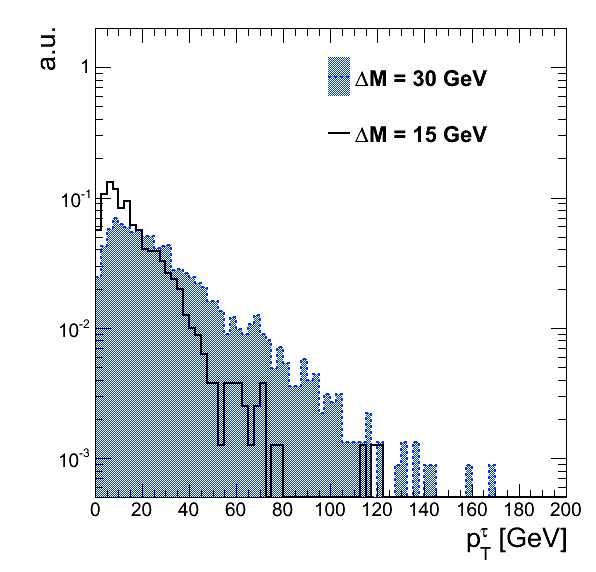}
\caption{$\pT$ of $\tau$ distribution normalized to arbitrary units in $\geq 2j + 2\tau$ final state for $ \Delta M = m_{\tilde{\chi}^{\pm}_1}-m_{\tilde{\chi}^0_1} = 30$ GeV and $15$ GeV.}
\label{VBFTauPt}
\end{figure}

In Figure \ref{VBFTauPt}, the normalized distribution of the $\pT$ of $\tau$ is displayed for $ \Delta M = m_{\tilde{\chi}^{\pm}_1}-m_{\tilde{\chi}^0_1} = 30$ GeV and $15$ GeV. For smaller $\Delta M$, the distribution peaks at lower $\pT$ and the signal acceptance is less efficient.

In \cite{Dutta:2012xe}, results were presented with $25$ fb$^{-1}$ of data at LHC8. We scale the results to LHC14, using the facts that $(i)$ the Wino production cross-section via VBF processes at the benchmark point is about twice at LHC14 compared to LHC8 and $(ii)$ the $VV+$ jets background that survives VBF cuts is about four times larger at LHC14 compared to LHC8. The significance $S/\sqrt{S+B}$ is thus approximately unchanged.

The $5 \sigma$ mass reach at $300$ fb$^{-1}$ ($1000$ fb$^{-1}$) of LHC14 for inclusive and opposite-sign $\tau$ pairs are $250$ GeV ($300$ GeV) and $200$ GeV ($250$ GeV), respectively, for $\Delta M \, = \, m_{\tilde{\tau}_1} - m_{\neu{1}} \, = \, 30$ GeV. The mass reach in the same-sign final state is similar to that in the opposite-sign case. For $\Delta M \, = \, 15$ GeV, the $3 \sigma$ mass reach at $300$ fb$^{-1}$ ($1000$ fb$^{-1}$) of LHC14 for inclusive $\tau$ pairs is $180$ GeV.

We have also considered the $t{\overline{t}}$ background and found that for $\geq 2j + 2\tau$ final state the VBF cut efficiency is $ 10^{-3}$, and the combined efficiency of $\met$, $2 \tau$ inclusive selection and  loose $b$-veto is $\sim 10^{-5}$, which renders it small compared to the other backgrounds. The VBF cuts are very effective in reducing the background in this case unlike the VBF production of Higgs ~\cite{cms1,atlas1} where much lower jet $\pT$  ($\pT>25$ GeV for ATLAS, $\pT > 30$ GeV for CMS) is used for VBF selection. Also, $M_{j_1j_2}$ is much smaller in the VBF production of Higgs (for example  CMS used $M_{j_1j_2} > 450$ GeV). In the SUSY case,  due to the requirement of larger energy  in the VBF system to produce chargino/neutralino pair, the jet $\pT$ is higher which is utilized to select the signal events~\cite{Choudhury:2003hq}.

\subsection{$\geq 2 j \, + \, 2 \mu \, + \, \met$}

In the previous section, the VBF search strategy in the multi-$\tau$ final state has been outlined. Although the search with $\tau$ leptons are expected to be the most daunting case amongst the three generation of leptons, it is worthwhile to study the potential sensitivity in cases where the charginos and neutralinos preferentially decay to the first two generation of sleptons via
\be
\neu{2} \, \rightarrow \, \tilde{l}^{\pm}_1 l^{\mp} \, \rightarrow \,  l^{\pm} l^{\mp} \neu{1} \,\,.
\ee

In this subsection, results for event selection with $\geq 2 j \, + \, 2 \mu \, + \, \met$ are presented. A branching ratio of $100\%$ of $\chpm{1}$ and $\neu{2}$ to $\tilde{\mu}$ has been assumed. After VBF selections, the following selections are employed:

$(i)$ Two isolated $\mu$'s with $\pT \geq 20$ GeV and $\pT \geq 15$ GeV in $|\eta| < 2.1$.

$(ii)$ $\met > 75$ GeV.  

$(iii)$ We also apply a loose $b$-veto which is useful in reducing the $t\overline{t}$ background.

The VBF cut efficiency for $t\overline{t}$ background is $10^{-3}$, and the combined efficiency of $\met$ cut, 2$\mu$ inclusive event selection, and loose $b$-veto is $\sim 2\times 10^{-4}$. The $t\overline{t}$ background is small compared to the other backgrounds. 

The $5 \sigma$ mass reach at $300$ fb$^{-1}$ ($1000$ fb$^{-1}$) of LHC14 for inclusive and opposite-sign $\mu$ pairs are approximately $350$ GeV ($400$ GeV) and $300$ GeV ($350$ GeV), respectively. The mass reach in the same-sign final state is similar to that in the opposite-sign case.

\section{Conclusion} \label{conclusion}

This whitepaper has investigated the direct production of $\neu{1}$ DM, as well as the production of $\chpm{1}$ and $\neu{2}$, by VBF processes at the LHC at $\sqrt{s} = 14$ TeV. The presence of high $E_T$ forward jets in opposite hemispheres with large dijet invariant mass is used to identify the VBF production. Kinematic requirements to search for signals of these supersymmetric particles above SM background arising from VBF and non-VBF processes have been developed. They have been shown to be effective in searching for direct production of $\neu{1}$ as well as the production of $\chpm{1}$ and $\neu{2}$ in both $2l$ as well as $2 \tau$ final states. 

For the DM study, it has been shown that broad enhancements in the $\met$ and VBF dijet mass distributions provide conclusive evidence for VBF production of supersymmetric DM. By optimizing the $\met$ cut for a given $m_{\neu{1}}$, one can simultaneously fit the $\met$ shape and observed rate in data to extract the mass and composition of $\neu{1}$, and hence solve for the DM relic density. At an integrated luminosity of $1000$ fb$^{-1}$, a significance of $5\sigma$ can be obtained up to a Wino mass of approximately $600$ GeV. The relic density can be determined to within $20\% \, (40\%)$ for the case of a pure Wino (Higgsino) for $500$ fb$^{-1}$ at LHC14, for $m_{\neu{1}} = 100$ GeV. We note that our study does not include the effect of large multiple interactions at high luminosity operations at the LHC. This is a very important subject, but outside the scope of the present work, because the final performance will depend on the planned upgrade of ATLAS and CMS detectors.

For the chargino-neutralino study, the $5 \sigma$ mass reach at $300$ fb$^{-1}$ ($1000$ fb$^{-1}$) of LHC14 for inclusive and opposite-sign $\tau$ pairs are $250$ GeV ($300$ GeV) and $200$ GeV ($250$ GeV), respectively, for $\Delta M \, = \, m_{\tilde{\tau}_1} - m_{\neu{1}} \, = \, 30$ GeV. The mass reach in the same-sign final state is similar to that in the opposite-sign case. For $\Delta M \, = \, 15$ GeV, the $3 \sigma$ mass reach at $300$ fb$^{-1}$ ($1000$ fb$^{-1}$) of LHC14 for inclusive $\tau$ pairs is $180$ GeV.

The $5 \sigma$ mass reach at $300$ fb$^{-1}$ ($1000$ fb$^{-1}$) of LHC14 for inclusive and opposite-sign $\mu$ pairs are approximately $350$ GeV ($400$ GeV) and $300$ GeV ($350$ GeV), respectively. The mass reach in the same-sign final state is similar to that in the opposite-sign case.

The next-to-leading order QCD corrections to the VBF electroweak
production cross sections
have not been considered. The inclusion of the K factor, which is very
modest for VBF production
($\sim$ 5\%), would improve the signal significance~\cite{kfactor}.

Searches for the EW sector in $\tau$ final states in Drell-Yan production face the challenge of controlling the level of backgrounds due to the larger $\tau$ misidentification rate as well as maintaining low enough $p_T$ thresholds for triggering. The VBF searches  are capable of reducing the background to manageable levels and thus probing multi-$\tau$ final states. 

With increasing instantaneous luminosity, both ATLAS and CMS experiments are raising their $p_T$  thresholds for triggering objects. The VBF trigger offers a promising route to probe supersymmetric production free from trigger bias. This is complementary to the existing LHC searches based on Drell-Yan production.  

\section{Acknowledgements}

This work is supported in part by the DOE grant DE-FG02-95ER40917, by the World Class University (WCU) project through the National Research Foundation (NRF) of Korea funded by the Ministry of Education, Science \& Technology (grant No. R32-2008-000-20001-0), and by the National Science Foundation grant  PHY-1206044. K.S. is supported by NASA Astrophysics Theory Grant NNH12ZDA001N.


\begin{thebibliography}{99}


\bibitem{LHCsquarkgluino} 
  ATLAS Collaboration,
  arXiv:1208.0949 [hep-ex].  
%
  ATLAS Collaboration,
  J. High Energy Phys. {\bf 07}, 167 (2012)  [arXiv:1206.1760 [hep-ex]].  
%
  CMS Collaboration,
arXiv:1207.1898 [hep-ex].  
%

\bibitem{LHCsquarkgluino20ifb} 
  ATLAS Collaboration,
ATLAS-CONF-2013-047.


\bibitem{stops}  
ATLAS Collaboration,
ATLAS-CONF-2012-166.
%
ATLAS Collaboration, 
ATLAS-CONF-2012-167.
%
  B.~Dutta, T.~Kamon, N.~Kolev, K.~Sinha and K.~Wang,
  ``Searching for Top Squarks at the LHC in Fully Hadronic Final State, Phys. Rev. D 86, 075004 (2012) [arXix:1207.1873 [hep-ph]].  
%
  T.~Plehn and M.~Spannowsky,
  ``Top Tagging,''
  arXiv:1112.4441 [hep-ph].
  T.~Plehn, M.~Spannowsky and M.~Takeuchi,
  ``Stop searches in 2012,''
  arXiv:1205.2696 [hep-ph].
  D.~E.~Kaplan, K.~Rehermann and D.~Stolarski,
  ``Searching for Direct Stop Production in Hadronic Top Data at the LHC,''
  arXiv:1205.5816 [hep-ph].
  D.~S.~M.~Alves, M.~R.~Buckley, P.~J.~Fox, J.~D.~Lykken and C.~-T.~Yu,
  ``Stops and MET: the shape of things to come,''
  arXiv:1205.5805 [hep-ph].
  T.~Han, R.~Mahbubani, D.~G.~E.~Walker and L.~-T.~Wang,
  ``Top Quark Pair plus Large Missing Energy at the LHC,''  JHEP {\bf 05}, 117 (2009)  [arXiv:0803.3820 [hep-ph]].  
  Y.~Bai, H.~-C.~Cheng, J.~Gallicchio and J.~Gu,
  ``Stop the Top Background of the Stop Search,''  JHEP {\bf 07}, 110 (2012)  [arXiv:1203.4813 [hep-ph]].  
%
ATLAS Collaboration, 
``Search for direct top squark pair production in final states with one isolated lepton, jets, and missing transverse momentum in $\sqrt{s} = 7$ TeV pp collisions using $4.7$ fb$^{-1}$ of ATLAS data," ATLAS-CONF-2012-073.


\bibitem{squarkheavy}
N.~Arkani-Hamed, A.~Gupta, D.~E.~Kaplan, N.~Weiner and T.~Zorawski,
arXiv:1212.6971 [hep-ph]. 
A.~Arvanitaki, N.~Craig, S.~Dimopoulos and G.~Villadoro,
arXiv:1210.0555 [hep-ph].



\bibitem{Arnowitt:2008bz} 
  R.~L.~Arnowitt, B.~Dutta, A.~Gurrola, T.~Kamon, A.~Krislock and D.~Toback,
  ``Determining the Dark Matter Relic Density in the mSUGRA Neutralino-Stau Co-Annhiliation Region at the LHC,''  Phys.\ Rev.\ Lett.\  {\bf 100}, 231802 (2008)  [arXiv:0802.2968 [hep-ph]].  

\bibitem{Hinchliffe:1996iu}
  I.~Hinchliffe, F.~E.~Paige, M.~D.~Shapiro, J.~Soderqvist and W.~Yao,
  ``Precision SUSY measurements at CERN LHC,''
  Phys.\ Rev.\  D {\bf 55}, 5520 (1997)
  [arXiv:hep-ph/9610544].
  I.~Hinchliffe and F.~E.~Paige,
  ``Measurements in SUGRA models with large tan(beta) at LHC,''
  Phys.\ Rev.\  D {\bf 61}, 095011 (2000)
  [arXiv:hep-ph/9907519].
  
\bibitem{Dutta:2011kp} 
  B.~Dutta, T.~Kamon, A.~Krislock, K.~Sinha and K.~Wang,
  ``Diagnosis of Supersymmetry Breaking Mediation Schemes by Mass Reconstruction at the LHC,''  Phys.\ Rev.\ D {\bf 85}, 115007 (2012)  [arXiv:1112.3966 [hep-ph]].  
  
\bibitem{ATLASneutchargino}
ATLAS Collaboration, 
ATLAS-CONF-2013-035;
ATLAS-COM-CONF-2013-042.  
  
\bibitem{CMSneutchargino}
CMS Collaboration, 
CMS PAS SUS-12-022.

\bibitem{Dutta:2012xe} 
  B.~Dutta, A.~Gurrola, W.~Johns, T.~Kamon, P.~Sheldon and K.~Sinha,
  Phys.\ Rev.\ D {\bf 87}, 035029 (2013)  arXiv:1210.0964 [hep-ph].
  
\bibitem{Delannoy:2013ata} 
  A.~G.~Delannoy, B.~Dutta, A.~Gurrola, W.~Johns, T.~Kamon, E.~Luiggi, A.~Melo and P.~Sheldon {\it et al.},
Physical Review Letters (in press).  

  
\bibitem{Cahn:1983ip} 
  R.~N.~Cahn and S.~Dawson,
  ``Production of Very Massive Higgs Bosons,''  Phys.\ Lett.\ B {\bf 136}, 196 (1984)  [Erratum-ibid.\ B {\bf 138}, 464 (1984)].  
  
\bibitem{Bjorken:1992er} 
  J.~D.~Bjorken,
  ``Rapidity gaps and jets as a new physics signature in very high-energy hadron hadron collisions,''  Phys.\ Rev.\ D {\bf 47}, 101 (1993).  

\bibitem{Rainwater:1998kj} 
  D.~L.~Rainwater, D.~Zeppenfeld and K.~Hagiwara,
  ``Searching for $H\to\tau^+\tau^-$ in weak boson fusion at the CERN LHC,''  Phys.\ Rev.\ D {\bf 59}, 014037 (1998)  [hep-ph/9808468].  

\bibitem{Choudhury:2003hq} 
  D.~Choudhury, A.~Datta, K.~Huitu, P.~Konar, S.~Moretti and B.~Mukhopadhyaya,
  ``Slepton production from gauge boson fusion,''  Phys.\ Rev.\ D {\bf 68}, 075007 (2003)  [hep-ph/0304192].  
  A.~Datta and K.~Huitu,
  ``Characteristic slepton signal in anomaly mediated SUSY breaking models via gauge boson fusion at the CERN LHC,''  Phys.\ Rev.\ D {\bf 67}, 115006 (2003)  [hep-ph/0211319].  
\bibitem{cho}G.~-C.~Cho, K.~Hagiwara, J.~Kanzaki, T.~Plehn, D.~Rainwater and T.~Stelzer,
  ``Weak boson fusion production of supersymmetric particles at the CERN LHC,''
  Phys.\ Rev.\ D {\bf 73}, 054002 (2006)
  [hep-ph/0601063].
\bibitem{datta}A.~Datta, P.~Konar and B.~Mukhopadhyaya,
  ``Signals of neutralinos and charginos from gauge boson fusion at the Large Hadron Collider,''
  Phys.\ Rev.\ D {\bf 65}, 055008 (2002) [hep-ph/0109071]; ``Invisible charginos and neutralinos from gauge boson fusion: A Way to explore anomaly mediation?,''
  Phys.\ Rev.\ Lett.\  {\bf 88}, 181802 (2002)
  [hep-ph/0111012]; P.~Konar and B.~Mukhopadhyaya,
  ``Gauge boson fusion as a probe of inverted hierarchies in supersymmetry,''
  Phys.\ Rev.\ D {\bf 70}, 115011 (2004)
  [hep-ph/0311347];
[hep-ph/0311347]; R.~C.~Cotta, J.~L.~Hewett, M.~P.~Le and T.~G.~Rizzo,
  arXiv:1210.0525 [hep-ph].

\bibitem{WMAP}
Planck Collaboration, arXiv:1303.5076 [astro-ph.CO]; 
WMAP Collaboration,   arXiv:1212.5226 [astro-ph.CO].





\bibitem{Giudice:2010wb} 
  G.~F.~Giudice, T.~Han, K.~Wang and L.~-T.~Wang,
  Phys.\ Rev.\ D {\bf 81}, 115011 (2010)  [arXiv:1004.4902 [hep-ph]].  



\bibitem{content}B.~Dutta, T.~Kamon, N.~Kolev, K.~Sinha, K.~Wang and S.~Wu,
  arXiv:1302.3231 [hep-ph].  
  


\bibitem{relic}R.~L.~Arnowitt, B.~Dutta, A.~Gurrola, T.~Kamon, A.~Krislock and D.~Toback,
  Phys.\ Rev.\ Lett.\  {\bf 100}, 231802 (2008);  B.~Dutta, A.~Gurrola, T.~Kamon, A.~Krislock, A.~B.~Lahanas, N.~E.~Mavromatos and D.~V.~Nanopoulos,
  Phys.\ Rev.\ D {\bf 79}, 055002 (2009); B.~Dutta, T.~Kamon, A.~Krislock, N.~Kolev and Y.~Oh,
  Phys.\ Rev.\ D {\bf 82}, 115009 (2010); B.~Dutta, T.~Kamon, A.~Krislock, K.~Sinha and K.~Wang,
  Phys.\ Rev.\ D {\bf 85}, 115007 (2012).  




\bibitem{Alwall:2011uj} 
  J.~Alwall, M.~Herquet, F.~Maltoni, O.~Mattelaer and T.~Stelzer,
  J. High Energy Phys. {\bf 06}, 128 (2011)  [arXiv:1106.0522 [hep-ph]].  



\bibitem{chargedtrack} 
  ATLAS Collaboration,
  J. High Energy Phys. {\bf 01}, 131 (2013)  [arXiv:1210.2852 [hep-ex]].  
%
  ALEPH Collaboration,
  Phys.\ Lett.\ B {\bf 533}, 223 (2002)  [hep-ex/0203020].  
  OPAL Collaboration,
  Eur.\ Phys.\ J.\ C {\bf 29}, 479 (2003)  [hep-ex/0210043].  
  DELPHI Collaboration,
  Eur.\ Phys.\ J.\ C {\bf 34}, 145 (2004)  [hep-ex/0403047].  

\bibitem{zeppenfeld} O.~J.~P.~Eboli and D.~Zeppenfeld,
  Phys.\ Lett.\ B {\bf 495}, 147 (2000)
  [hep-ph/0009158].

\bibitem{Chatrchyan:2012jua} 
  CMS Collaboration,
  JINST {\bf 8}, P04013 (2013)  [arXiv:1211.4462 [hep-ex]].  

\bibitem{Allahverdi:2012wb} 
  R.~Allahverdi, B.~Dutta and K.~Sinha,
  Phys.\ Rev.\ D {\bf 86}, 095016 (2012)  [arXiv:1208.0115 [hep-ph]].  
  H.~Baer, V.~Barger and D.~Mickelson,
  arXiv:1303.3816 [hep-ph].  


\bibitem{Allahverdi:2012gk} 
  R.~Allahverdi, B.~Dutta and K.~Sinha,
  arXiv:1212.6948 [hep-ph].  

\bibitem{Baer:2012cf} 
  H.~Baer, V.~Barger, P.~Huang, D.~Mickelson, A.~Mustafayev and X.~Tata,
  arXiv:1212.2655 [hep-ph].  


\bibitem{Griest:1990kh} 
  K.~Griest and D.~Seckel,
  ``Three exceptions in the calculation of relic abundances,''  Phys.\ Rev.\ D {\bf 43}, 3191 (1991).  
  
\bibitem{Carena:2011aa} 
  M.~Carena, S.~Gori, N.~R.~Shah and C.~E.~M.~Wagner,
  ``A 125 GeV SM-like Higgs in the MSSM and the $\gamma \gamma$ rate,''  JHEP {\bf 03}, 014 (2012)  [arXiv:1112.3336 [hep-ph]].  


\bibitem{Alwall:2011uj} 
J.~Alwall, ``MadGraph/MadEvent v4: the new web generation",
JHEP {\bf 09}, 028 (2008)  [arXiv:0706.2334  [hep-ph]].


\bibitem{pythia}
T. Sjostrand, S. Mrenna, and P. Skands,
 "PYTHIA 6.4 Physics and Manual."
  J. High Energy Phys. \textbf{05} (2006) 026.

\bibitem{pgs}
\pgs\ is a parameterized detector simulator.
We use version 4
(\url{http://www.physics.ucdavis.edu/~conway/research/software/pgs/pgs4-general.htm})
in the LHC detector configuration.
The $b$-jet tagging efficiency in PGS is
$\sim$42\% for $E_{\rm T} >$ 50 GeV
and $|\eta| < 1.0$, and degrading between  $1.0 < |\eta| < 1.5$.
The $b$-tagging fake rate for $c$ and light quarks/gluons is
$\sim9\%$ and $2\%$, respectively.
So far,  no problem has been reported to carry out 8 TeV in any analyses.

 \bibitem{CMS-PAS-FSQ-12-019}CMS Collaboration, ``Measurement of the electroweak production cross section of the $Z$ boson with two forward-backward jets in pp collisions at 7 TeV”, CMS-PAS-FSQ-12-019.

\bibitem{atlaspaper}ATLAS collaboration, ``Jet energy measurement with the ATLAS detector in proton-proton collisions at $\sqrt{s}=7$ TeV,'' arXiv:1112.6426.

\bibitem{cmstau}
CMS Collaboration, ``Performance of $\tau$-lepton reconstruction and identification in CMS”, 
J. Instrum. 07 P01001 (2012), doi:10.1088/1748-0221/7/01/P01001.

\bibitem{cms1}
CMS Collaboration, ``Search for the standard model Higgs boson decaying to $W^+ W^−$ in the fully leptonic final state in $pp$ collisions at $\sqrt{s} = 8$ TeV," CMS-PAS-HIG-12-038.

\bibitem{atlas1}
ATLAS Collaboration, ``Search for the Standard Model Higgs boson in the $H \rightarrow$ WW(*) $\rightarrow \ell \nu \ell \nu$ decay mode with $4.7$ fb$^{-1}$ of ATLAS data at $\sqrt{s}=7$ TeV,''  Phys.\ Lett.\ B {\bf 716}, 62 (2012)  [arXiv:1206.0756 [hep-ex]].  
\bibitem{kfactor}T.~Figy, C.~Oleari and D.~Zeppenfeld,
  ``Next-to-leading order jet distributions for Higgs boson production via weak boson fusion,''
  Phys.\ Rev.\ D {\bf 68}, 073005 (2003)
  [hep-ph/0306109]; P.~Konar and D.~Zeppenfeld,
  ``Next-to-leading order QCD corrections to slepton pair production via vector-boson fusion,''
  Phys.\ Lett.\ B {\bf 647}, 460 (2007)
  [hep-ph/0612119]; C.~Oleari and D.~Zeppenfeld,
  Phys.\ Rev.\ D {\bf 69}, 093004 (2004)
  [hep-ph/0310156].




\end{thebibliography}
\end{document}